\documentclass[12pt,a4paper]{article}
\usepackage[pdfstartview=FitH,colorlinks=true,linkcolor=black,anchorcolor=black,citecolor=black,urlcolor=black]{hyperref}
\usepackage[english]{babel}
\usepackage{amsmath,amssymb,titling,authblk}
\usepackage{slashed}
\usepackage{amsmath}
\usepackage{amscd}
\usepackage{mathrsfs}
\usepackage[normalem]{ulem}
\usepackage{appendix}
\usepackage{cite}

\ifx\pdfoutput\undefined
\usepackage{graphicx}
\else
\usepackage[pdftex]{graphicx}
\fi

\begin{document}

\author[1,2]{Oleg O.~Novikov\thanks{o.novikov@spbu.ru}}

\author[3,4]{Arina M.~Shtennikova\thanks{arshtm@gmail.com}}

\affil[1]{Saint Petersburg State University, 7/9 Universitetskaya nab.\\
St. Petersburg, Russia 199034}

\affil[2]{Immanuel Kant Baltic Federal University,
Al.Nevsky St. 14, Kaliningrad, 236041, Russia}

\affil[3]{Department of Particle Physics and Cosmology, Physics Faculty,M.V. Lomonosov Moscow State University,Vorobjevy Gory, 119991 Moscow, Russia}

\affil[4]{ Institute for Nuclear Research of the Russian Academy of Sciences,60th October Anniversary Prospect, 7a, 117312 Moscow, Russia}

\title{Quantum decoherence of the homogeneous modes during inflation}

\maketitle

\begin{abstract}
In this short note we consider the decoherence of the minisuperspace background degrees of freedom due to the long wavelength scalar and tensor perturbations becoming unobservable within the Born-Oppenheimer approach to the quantum gravity. We apply this to the power law inflation model.
\end{abstract}

\section{Introduction}

According to the observations the Universe is spatially flat, very homogeneous and isotropic \cite{Planck2018}. To account for these facts within the Hot Big Bang model would require a high degree of fine-tuning. However, they are successfully explained in the cosmological inflation paradigm \cite{gorb,B}. It states that early in the cosmological evolution the Universe experienced an exponentially fast expansion. If this is true, the whole observable universe originates from a tiny region that explains a high degree of homogeneity. The cosmological inflation theory also successfully predicts the spectrum of the primordial fluctuations. They are assumed to originate from the quantum fluctuations of the primordial vacuum state that was transformed into a squeezed state by the cosmological expansion \cite{Muh}.

The state in the beginning of the cosmological evolution is essentially non-classical.
This raises the question how the transition to the observed classicalized state occurs. The problem of the quantum-to-classical transition also arises for the mesoscale systems. In that context the crucial role is played by decoherence processes due to the interaction with environment \cite{Sch}. The interaction leads to the entanglement between the system and environmental degrees of freedom. As result for the measurements of the sufficiently coarse-grained observables of the system the quantum interference becomes strongly suppressed and it may demonstrate the effectively classical random behavior.

Within the cosmological context the transition of the inhomogeneities to the classical behavior was considered within the quantum field theory on the curved spacetime background. The classicalization of the fluctuations may happen due to the squeezing of the quantum state during the cosmological expansion and due to the loss of the information about the long wavelength modes \cite{dec1,dec2,dec3}. Another aspect of the problem arises when the homogeneous background modes are included into consideration as is done within the Wheeler-DeWitt approach \cite{kief}. The quantum state is, generally speaking, a wave packet spread over various background configurations. Hence the question of the classicalization may be posed for the large-scale degrees of freedom that serve as a background for the small-scale inhomogeneities. In \cite{qc1} the example of the de Sitter spacetime was considered with the decoherence happening due to the loss of information about long wavelength modes beyond the cosmological horizon. In contrast, the works \cite{qc2,qc3,qc4,qc5} use the coarse-graining approach associating the classicalization with lack of information about short wavelength modes including the gravitational waves.

In this short note we further study the decoherence of the background degrees of freedom caused by the loss of the information about the large wavelength modes due to the appearance of the cosmological horizon, applying it to a power law inflation.

\section{Wheeler-DeWitt equation}

Consider the cosmology of a scalar field minimally coupled to the gravity,
\begin{equation}
S=\int d^4x \Big[\frac{1}{2}R+\frac{1}{2}g^{\mu\nu}\partial_\mu\Phi\partial_\nu\Phi-V(\Phi)\Big]
\end{equation}
The cosmological solution may be represented as the approximately flat Friedmann-Robertson-Walker cosmology with inhomogeneous perturbations,
\begin{align}
ds^2=&N(\eta)^2 e^{2\alpha(\eta)}\Big(1-2S(\eta,\vec{x})\Big)d\eta^2-2 N(\eta) e^{2\alpha(\eta)}\Big(V_k(\eta,\vec{x})+\partial_k V(\eta,\vec{x})\Big)d\eta\, dx^k\nonumber\\
&-e^{2\alpha(\eta)}\Bigg[\Big(1-2\psi(\eta,\vec{x})\Big)\delta_{ij}+\partial_i F_j(\eta,\vec{x})+\partial_j F_i(\eta,\vec{x}) + 2\partial_i\partial_j E(\eta,\vec{x})\Bigg]dx^i\, dx^j\nonumber\\
&-e^{2\alpha(\eta)}b_{ij} dx^i\,dx^j,
\end{align}
\begin{equation}
\Phi(\eta,\vec{x})=\bar{\Phi}(\eta)+\phi(\eta,\vec{x})
\end{equation}
Up to the second order in the inhomogeneous perturbations the scalar ($S$, $V$, $\psi$, $E$, $\phi$), vector ($V_k$, $F_k$) and tensor ($b_{ij}$) variables are separated from each other. Because the vector sector yields only the rapidly decreasing solutions, we will neglect such perturbations in this work.

Then up to the second order in the inhomogeneous perturbations the Hamiltonian constraint takes the form,
\begin{equation}
\mathcal{H}=\mathcal{H}_0+\mathcal{H}_S+\mathcal{H}_T
\end{equation}

Here $\mathcal{H}_0$ denotes the Hamiltonian for the homogeneous modes. To properly define it we must introduce a certain infrared regularization with a finite spatial volume $\mathcal{L}^3$. We will assume that this spatial volume (measured in the Planck units) is sufficiently large so that the details of the regularization should not affect the result. The homogeneous Hamiltonian then takes the form,
\begin{equation}
\mathcal{H}_0=e^{-2\alpha}\Big[-\frac{1}{12\mathcal{L}^3}\pi_{\alpha}^2+\frac{1}{2\mathcal{L}^3}\pi_{\bar{\Phi}}^2+\mathcal{L}^3 V(\bar{\Phi})\Big],
\end{equation}
where $\pi_\alpha$ and $\pi_{\bar{\Phi}}$ are the canonical momenta for $\alpha$ and $\bar{\Phi}$ variables.

The scalar sector can be studied in terms of the Mukhanov-Sasaki variable $v$. After the infrared regularization the inhomogeneities may be decomposed into a discrete spectrum of the Laplace operator eigenvalues,
\begin{equation}
v(\eta,\vec{x})=\sum_n v_n(\eta) f_n(\vec{x}),\quad \Delta f_n(\vec{x})=-\omega_n^2 f_n(\vec{x})
\end{equation}
Similarly the tensor perturbations can be decomposed into a sum over polarizations,
\begin{equation}
b_{ij}(\eta,\vec{x})=\sum_{s=+,\times}\sum_n b_{s,n}(\eta) g_n^{(s)}(\vec{x}),\quad \Delta g_n^{(s)}(\vec{x})
\end{equation}

Then the scalar and tensor parts of the Hamiltonian constraint take the form,
\begin{equation}
\mathcal{H}_S=\sum_n \Bigg[\frac{1}{2}\pi_{v,n}^2+\frac{1}{2}\omega_{S,n}^2 v_{n}^2\Bigg],\quad \omega_{S,n}^2=\omega_n^2-\frac{z''}{z}
\end{equation}
\begin{equation}
\mathcal{H}_T=\sum_{s=+,\times}\sum_n \Bigg[\frac{1}{2}\pi_{b,s,n}^2+\frac{1}{2}\omega_{T,n}^2 b_{s,n}^2\Bigg],\quad \omega_{T,n}^2=\omega_n^2-(\alpha')^2-\alpha''
\end{equation}

The Wheeler-DeWitt equation for the wave function of the Universe is,
\begin{equation}
\Big[\mathcal{H}_0+\mathcal{H}_S+\mathcal{H}_T\Big]\Psi\Big(\alpha,\bar{\Phi},\{v_n,b_{+,n},b_{\times,n}\}\Big)=0
\end{equation}

We will employ the Born-Oppenheimer approximation \cite{kief,BOKam,BOKam1}, that combined with the separation of variables for the inhomogeneous perturbations yields,
\begin{align}
\Psi\Big(\alpha,\bar{\Phi},\{v_n,b_{+,n},b_{\times,n}\}\Big)\simeq \Psi_0(\alpha,\bar{\Phi})\nonumber\\
\cdot\prod_{n}\psi_{S,n}(\{v_n\}|\alpha,\bar{\Phi})\psi_{T,n}^{(+)}(\{b_{+,n}\}|\alpha,\bar{\Phi})\psi_{T,n}^{(\times)}(\{b_{\times,n}\}|\alpha,\bar{\Phi})
\end{align}
Let us denote $Q_a=\{\alpha,\bar{\Phi}\}$ and $q_{I,n}=\{v_n,b_{+,n},b_{\times,n}\}$.

Let us assume that the homogeneous wave function can be described by the WKB-wavepacket,
\begin{equation}
\Psi_0(\alpha,\bar{\Phi})\simeq \mathcal{A}(\alpha,\bar{\Phi})\exp\Big\{i\mathcal{L}^3\mathcal{S}(\alpha,\bar{\Phi})-\mathcal{L}^3\mathcal{R}(\alpha,\bar{\Phi})\Big\}
\end{equation}
where $\mathcal{R}$ provides the shape of the wavepacket. We will look at the region near its center where $|\nabla\mathcal{R}|\ll|\nabla\mathcal{S}|$.

The minisuperspace Wheeler-DeWitt equation admits the conserved current,
\begin{equation}
J_a=|\mathcal{A}|^2\partial_a\mathcal{S},\quad -\frac{1}{6}\partial_\alpha J_\alpha+\partial_{\bar{\Phi}}J_{\bar{\Phi}}
\end{equation}
Let us also introduce the derivative in the so-called WKB-time \cite{Brout1,Brout2},
\begin{equation}
\frac{\partial}{\partial\eta}=e^{-2\alpha}\Big[-\frac{1}{6}\partial_\alpha\mathcal{S}\partial_\alpha+\partial_{\bar{\Phi}}\mathcal{S}\partial_{\bar{\Phi}}\Big]
\end{equation}
This derivative is orthogonal to the surfaces of $\mathcal{S}=\mathrm{const}$ and is directed along the classical trajectories. We can approximately identify it with the conformal time on the classical solution. This remind the equation for the cosmological fluctuations on the fixed background \cite{deruelle}. Then from the Wheeler-DeWitt equation it follows,
\begin{align}
i\frac{\partial}{\partial\eta}\psi_{I,n}(q_{I,n}|\alpha,\bar{\Phi})=\Big[-\frac{1}{2}\frac{\partial^2}{\partial q_{I,n}^2}+\frac{1}{2}\omega_{I,n}^2(\eta) q_{I,n}^2\Big]\psi_{I,n}(q_{I,n}|\alpha,\bar{\Phi})
\end{align}
where $\omega_{I,n}^2$ is computed on a classical trajectory that passes through $(\alpha,\bar{\Phi})$ in the direction $\frac{\partial}{\partial\eta}$. It admits the conserved inner product,
\begin{equation}
\langle \psi_{I,n}^{(1)}|\psi_{I,n}^{(2)}\rangle_{(\alpha,\bar{\Phi})}=\int_{-\infty}^{+\infty}dq_{I,n}\,\Big(\psi_{I,n}^{(1)}(q_{I,n}|\alpha,\bar{\Phi})\Big)^\ast\psi_{I,n}^{(2)}(q_{I,n}|\alpha,\bar{\Phi})
\end{equation}
Then for two wave packets with the same $\mathcal{S}$ but different $\mathcal{A}$ defined on a surface $\Sigma$ of $\mathcal{S}=\mathrm{const}$ the following inner product is approximately conserved with evolution of $\mathcal{\Sigma}$ in the WKB-time \cite{Novikov:2017dsl},
\begin{equation}
\langle \Psi^{(1)}|\Psi^{(2)}\rangle_{\Sigma}=\int_\Sigma d\Sigma \Big(\mathcal{A}^{(1)}(\alpha,\bar{\Phi})\Big)^\ast\mathcal{A}^{(2)}(\alpha,\bar{\Phi})(\partial_\perp\mathcal{S})\prod_{I,n}\langle \psi_{I,n}^{(1)}|\psi_{I,n}^{(2)}\rangle_{(\alpha,\bar{\Phi})}
\end{equation}
Thus $\mathcal{A}\psi_S\psi_T^{(+)}\psi_T^{(\times)}$ can be treated as a WKB-time dependent quantum state defined on $\Sigma$.

\section{Decoherence of the background modes}

The inhomogeneous mode state is assumed to be a vaccuum state at the start of the inflation. Such state should be annihilated by a certain annihilation operator,
\begin{equation}
\hat{a}_{I,n}\psi_{I,n}=0,\quad \hat{a}_{I,n}=if^\ast\hat{\pi}_{I,n}-i(f')^\ast q_{I,n}
\end{equation}
where $f$ is a certain solution of the classical equation of motion,
\begin{equation}
f''+\omega_{I,n}^2(\eta)f=0,\quad f\underset{\eta\rightarrow-\infty}{\longrightarrow}\frac{1}{\sqrt{2\omega_{n}}}\exp\Big(-i\omega_{n}\eta\Big)
\end{equation}
We may find it as,
\begin{equation}
\psi_{I,n}(q_{I,n})=\frac{1}{(2\pi)^{1/4}}\frac{1}{\sqrt{f^\ast}}\exp\Big(\frac{i}{2}\frac{(f')^\ast}{f^\ast}q_{I,n}^2\Big)
\end{equation}

Let us consider the power-law inflation $e^{\alpha}\sim t^p$ with slow-roll condition, i.e. $\dot{\Phi}\ll V(\Phi),\; \ddot{\Phi}\ll 3H\dot{\Phi}$. In this case the slow-roll parameters can be expressed exactly,
\begin{equation}
\varepsilon = -\frac{\dot{H}}{h^{2}}=\frac{1}{p},\quad \delta = -\frac{\ddot{\Phi}}{H\dot{\Phi}}=\frac{1}{p}
\end{equation}
The scaling parameter can be expressed in terms of the conformal time as $e^{\alpha} \sim \eta^{1+\beta}$. Then, the frequencies in the equation for the fluctuations can be rewritten in the following way,
\begin{equation}
\omega_{S,k}^{2} = \omega_{T,s,k}^{2} = \omega_n^2- \frac{2+3\varepsilon}{\eta^{2}} = \omega_n^2-\frac{\beta(1+\beta)}{\eta^{2}}, \qquad \text{where} \; 1+\beta = \frac{p}{1+p}
\end{equation}

Let's denote $\omega_n=k^2$. The equation for classical fluctuations,
\begin{equation}
\ddot{f}_{\vec{k}}+\left(\vec{k}^2-\frac{\beta(1-\beta)}{\eta^2}\right)f_{\vec{k}}=0
\end{equation}
is the Bessel equation. Its solutions can be written in terms of Hankel functions,
\begin{equation}
f_{k}= C_1\sqrt{\eta}H^{(1)}_{\beta+\frac{1}{2}}(k\eta)+C_2\sqrt{\eta}H^{(2)}_{\beta+\frac{1}{2}}(k\eta)
\end{equation}
Hankel functions have different asymptotical behavior at infinity \cite{NIST},
\begin{align}
H^{(1)}_{\nu}(z)\xrightarrow[\lvert z\rvert \rightarrow\infty]{} \sqrt{\frac{2}{\pi z}}e^{i(z-\frac{1}{2}\pi\nu-\frac{1}{4}\pi)}\\
H^{(2)}_{\nu}(z)\xrightarrow[\lvert z\rvert \rightarrow\infty]{} \sqrt{\frac{2}{\pi z}}e^{-i(z-\frac{1}{2}\pi\nu-\frac{1}{4}\pi)}
\end{align}
Therefore, to satisfy the asymptotic condition on $f_{k}$, one should choose a solution containing only the Hankel function of the first kind $H^{(1)}_{\beta+\frac{1}{2}}(k\eta)$. The properly normalize classical solution is,
\begin{eqnarray}
f_{k} = \frac{\sqrt{\pi}}{2}\sqrt{\eta}H^{(1)}_{\beta+\frac{1}{2}}(k\eta) \nonumber\\
f^{*}_{k} = \frac{\sqrt{\pi}}{2}\sqrt{\eta}H^{(2)}_{\beta+\frac{1}{2}}(k\eta)
\end{eqnarray}

Various classical solutions are parametrized by $\beta$, which is defined by the initial conditions. Lets recall that the wave function is a WKB wave-packet that can be considered as a bundle of classical trajectories, i.e.
\begin{equation}
\Psi\rvert_{\mathcal{S}=const}= \mathcal{A}(\beta)\prod_{n}\frac{1}{(2\pi)^{\frac{1}{2}}}\frac{1}{f^{*}}e^{\frac{i}{2}\frac{\dot{f}^{*}}{f^{*}}\left(v_{n}^{2}+\left(b_{n}^{(+)}\right)^{2}+\left(b_{n}^{(\times)}\right)^{2}\right)}
\end{equation}

The density matrix of the total state is,
\begin{equation}
\rho=\lvert\Psi\rangle\langle \Psi\rvert = \Psi(Q,\{q\})\Psi^{*}(\tilde{Q},\{q\})
\end{equation}

In our case, the density matrix takes the following form,

\begin{eqnarray}
&\rho(\beta,\tilde{\beta},\{v_{n},b_{T,+,n},b_{T,\times,n}\},\{v_{n},b_{T,+,n},b_{T,\times,n}\})= \nonumber\\
& \mathcal{A}(\beta)\mathcal{A}^*(\tilde{\beta})\prod_{n,m}\frac{1}{2\pi}\frac{1}{f^{*}(\beta)f(\tilde{\beta})}e^{\frac{i}{2}\frac{\dot{f}^{*}(\beta)}{f^{*}(\beta)}\left(v_{n}^{2}+\left(b_{n}^{(+)}\right)^{2}+\left(b_{n}^{(\times)}\right)^{2}\right)-\frac{i}{2}\frac{\dot{f}(\tilde{\beta})}{f(\tilde{\beta})}\left(v_{m}^{2}+\left(b_{m}^{(+)}\right)^{2}+\left(b_{m}^{(\times)}\right)^{2}\right)}
\end{eqnarray}

We now take into account the fact that conformal time coincides with the position of  cosmological horizon that defines a causally-connected region. If the wavelength of the perturbation is larger than the size of the horizon, then we cannot observe it in any way. Therefore, we are going to assume that these modes are a kind of environment for all the others, and consider a reduced density matrix,
\begin{eqnarray}
\rho^{red}=\int dv_{q}\, db_{q}^{(+)}\, db_{q}^{(\times)} \, \rho(\beta,\tilde{\beta},\{v_{n},b_{n}^{(+)},b_{n}^{(-)}\},\{\tilde{v}_{m},\tilde{b}_{m}^{(+)},\tilde{b}_{m}^{(\times)}\}),
\end{eqnarray}

where $q$ designates all $k<\frac{1}{\lvert\eta\rvert}$. Since this is Gauss integral, the reduced density matrix is,
\begin{eqnarray}
\rho^{red}&=&A(\beta)A^*(\tilde{\beta})\prod_{k<\frac{1}{\lvert\eta\rvert}} \left( \frac{i}{\dot{f}^*(\beta)f(\tilde{\beta})-\dot{f}(\tilde{\beta})f^*(\beta)}\right)^{\frac{3}{2}}\times \nonumber\\
	&\times&\prod_{k>\frac{1}{\lvert\eta\rvert}} \frac{1}{2\pi}\frac{1}{f(\tilde{\beta})f^*(\beta)}e^{-\frac{i}{2}\left( \frac{\dot{f}}{f}-\frac{\dot{f}^*}{f^*}\right)\left(v^2_{k}+\left(b_{k}^{(+)}\right)^{2}+\left(b_{T,k}^{(\times)}\right)^{2}\right)}
\end{eqnarray}

Let us separately analyze the part corresponding to the long-wavelength range, that is, one where $k<\frac{1}{\lvert\eta\rvert}$. Until now, we have been considering the discrete spectrum, but after turning off the infrared regularization it becomes a continuous one. To do this, we rewrite the first product in the following way,
\begin{eqnarray}
\prod_{k<\frac{1}{\lvert\eta\rvert}} \left( \frac{i}{\dot{f}^*(\beta)f(\tilde{\beta})-\dot{f}(\tilde{\beta})f^*(\beta)}\right)^{\frac{3}{2}}
&=& e^{\underset{k<\frac{1}{\lvert\eta\rvert}}{\sum}\frac{3}{2}\ln\left( \frac{i}{\dot{f}^*(\beta)f(\tilde{\beta})-\dot{f}(\tilde{\beta})f^*(\beta)}\right)} =\nonumber\\
&=& e^{\underset{k<\frac{1}{\lvert\eta\rvert}}{\sum}-\frac{3}{2}\ln \left((-i)\dot{f}^*(\beta)f(\tilde{\beta})-\dot{f}(\tilde{\beta})f^*(\beta)\right)}
\end{eqnarray}

The sum in the exponent can be replaced back by the integral,
\begin{equation}
\sum_{k<\frac{1}{\lvert\eta\rvert}} \{ \ldots\} \rightarrow 	\mathscr{L}^{3} \int_{\frac{1}{\eta}}^{0} d^{3}k \{ \ldots\}
\end{equation}

Let us write separately the denominator in the logarithm,
\begin{eqnarray}
\dot{f}^*(\beta)f(\tilde{\beta})-\dot{f}(\tilde{\beta})f^*(\beta) &=&\frac{k\eta\pi}{4}\left( H^{(1)}_{\tilde{\beta}+\frac{3}{2}}H^{(2)}_{\beta+\frac{1}{2}}+H^{(1)}_{\tilde{\beta}+\frac{1}{2}}H^{(2)}_{\beta+\frac{3}{2}}\right)+\nonumber\\
&+&\frac{\pi}{4}H^{(1)}_{\tilde{\beta}+\frac{1}{2}}H^{(2)}_{\beta+\frac{1}{2}}(\beta-\tilde{\beta})
\end{eqnarray}

Hankel functions represent a combination of Bessel functions of the first and second kind, and can be represented as a sum of two series,
\begin{eqnarray}
H^{(1)}_{\nu}(z)&=&\frac{i}{sin(\pi\nu)}\left(e^{-\nu\pi z}J_{\nu}(z)-J_{-\nu}(z)\right)\nonumber\\
H^{(2)}_{\nu}(z)&=&\frac{i}{sin(\pi\nu)}\left(J_{-\nu}(z)-e^{\nu\pi z}J_{\nu}(z)\right)\nonumber\\
J_{\nu}(z)&=&\left(\frac{1}{2}z\right)^{\nu}\sum_{k=0}^{\infty}\frac{(-1)^{\nu}\left(\frac{1}{4}z^{2}\right)^{k}}{k!\Gamma(\nu+k+1)}
\end{eqnarray}

Note that the integration interval is very small and close to zero. This means that the first terms of the series determine the behavior of the Hankel functions. Then,
\begin{eqnarray}
&-&i\left(\dot{f}^*(\beta)f(\tilde{\beta})-\dot{f}(\tilde{\beta})f^*(\beta)\right) = \nonumber \\
&=&\frac{\pi}{4}\,{\frac {(\tilde{\beta}+\beta+1)\left(\frac{1}{2}k\eta\right)^{\tilde{\beta}-\beta}e^{-i\pi\tilde{\beta}}}{\cos\left( \tilde{\beta}\,\pi  \right) \cos \left( \beta\,\pi\right)\left(\Gamma\left(\tilde{\beta}+\frac{1}{2} \right)\tilde{\beta}+\frac{1}{2}\,\Gamma\left(\tilde{\beta}+\frac{1}{2}\right)\right)\Gamma\left(\frac{1}{2}-\beta\right)}}+ \nonumber \\
&+&\frac{\pi}{4}\,{\frac {(\tilde{\beta}+\beta+1)\left(\frac{1}{2}k\eta\right)^{\beta-\tilde{\beta}}e^{i\pi\beta}}{\cos\left(\tilde{\beta}\,\pi\right)\cos\left(\beta\,\pi\right)\Gamma\left( \frac{1}{2} - \tilde{\beta} \right)\left(\Gamma\left(\beta+\frac{1}{2}\right)\beta+\frac{1}{2}\,\Gamma\left(\beta+\frac{1}{2}\right)\right)}}- \nonumber \\
&-&\frac{i\pi}{4}{\frac {(\beta-\tilde{\beta})\left(\frac{1}{2}k\eta\right)^{-1-\tilde{\beta}-\beta}}{\cos\left(\tilde{\beta}\,\pi\right) \cos \left(\beta\,\pi\right)\Gamma\left(\frac{1}{2}-\tilde{\beta}\right)\Gamma\left(\frac{1}{2}-\beta\right)}}+\nonumber \\
&+&\frac{i\pi}{4}\frac {\left(\frac{1}{2}k\eta\right)^{1+\tilde{\beta}+\beta}{e^{i\pi \, \left(\beta-\tilde{\beta} \right) }} \left( \beta-\tilde{\beta} \right) }{\cos \left( \beta\,\pi\right) \cos \left( \tilde{\beta}\,\pi  \right) \Gamma  \left( \tilde{\beta}+\frac{3}{2} \right) \Gamma  \left( \beta+\frac{3}{2} \right)}- \nonumber\\
&-&\frac{i\pi}{4}\frac{\left(\frac{1}{2}k\eta\right)^{1-\tilde{\beta}-\beta} \left( {\beta}^{2}-{\tilde{\beta}}^{2}+\beta-\tilde{\beta}\right)}{\cos\left(\tilde{\beta}\,\pi\right)\cos\left(\beta\,\pi\right)\Gamma\left(\frac{3}{2}-\beta\right)\Gamma\left(\frac{3}{2}- \tilde{\beta}\right)}+\nonumber \\
&+&\frac{i\pi}{2}\,\frac{\left(\frac{1}{2}k\eta\right)^{\tilde{\beta}+\beta+3}{e^{i\pi\,\left(\beta-\tilde{\beta}\right)}}\left(\beta-\tilde{\beta}\right)}{\cos\left(\beta\,\pi\right)\cos\left(\tilde{\beta}\,\pi\right)\Gamma\left(\beta+5/2\right)\Gamma\left(\tilde{\beta}+5/2\right)}- \nonumber \\
&-&\frac{i\pi}{4}\,\frac {\left(\frac{1}{2}k\eta\right)^{3-\tilde{\beta}-\beta} \left(\beta-\tilde{\beta}\right)}{\cos\left(\tilde{\beta}\,\pi\right)\cos\left(\beta\,\pi\right)\Gamma\left(\frac{3}{2}-\beta\right)\Gamma\left(\frac{3}{2}- \tilde{\beta}\right)}
\end{eqnarray}

The first two terms in the line $\beta = \tilde{\beta}$ are equal to one, and the rest to zero, it indicates that the normalization for the density matrix is conserved. Furthermore, for different $\beta$ and $\tilde{\beta}$ the expression module becomes less than 1 in the range $(\beta-\tilde{\beta})\in (-\frac{1}{2},\frac{1}{2})$ and $(\beta+\tilde{\beta})\in (-1,0)$. Since different $\beta$ correspond to off-diagonal terms of the reduced density matrix, we observe their suppression in comparison to the main diagonal.

Due to the complexity of the integrand, the integration was performed numerically. As a result, we obtained the following graphs of the short-wave part of the density matrix for different values $\eta$,

\begin{figure}[h!]
\begin{minipage}[h!]{0.5\linewidth}
\center{\includegraphics[width=0.9\linewidth]{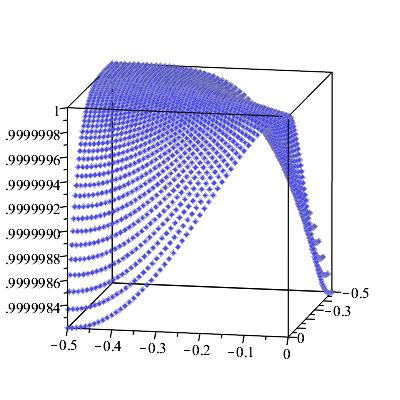} \\ $\eta=-10$}
\end{minipage}
\hfill
\begin{minipage}[h!]{0.49\linewidth}
\center{\includegraphics[width=0.9\linewidth]{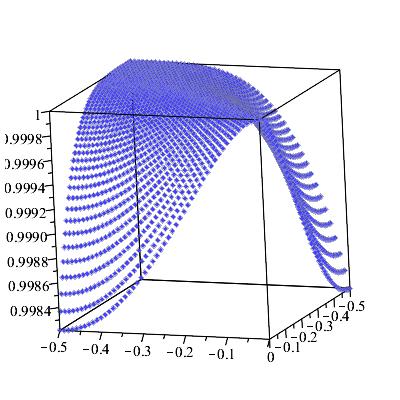} \\ $\eta=-1$}
\end{minipage}
\hfill
\begin{minipage}[h!]{0.49\linewidth}
\center{\includegraphics[width=0.9\linewidth]{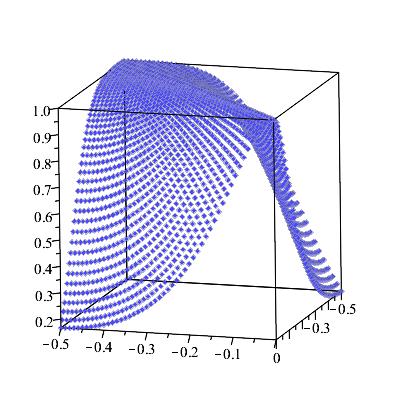} \\ $\eta=-0.1$}
\end{minipage}
\end{figure}

The graphs clearly show the properties discussed above. The diagonal is normalized to 1; as the difference between $\beta-\tilde{\beta}$ increases, the off-diagonal terms decrease. With a decrease in $|\eta|$, that is with cosmic time $t\rightarrow+ \infty$, the diagonal elements dominate, so the system behaves in an increasingly classical way, and the interference is suppressed.

\section{Conclusion}

We considered the evolution of the universe, initially in a vacuum state, in the period of inflation. Over time, an increasing number of perturbation modes of the metric and inflaton field goes beyond the cosmological horizon and becomes unobservable. It was shown that during this process the density matrix becomes mixed for short-wave modes, in the limit diagonal in the $\beta$ variable characterizing the rate of inflation. This suggests that from the point of view of the local observer the coherent wave package in late times corresponds to the classical probability distribution of background metrics.

\section*{Acknowledgements}
The work is supported by the Program of development of the regional scientific and educational mathematical center "North-western center of the mathematical studies named by Sofia Kovalevskaya" from the Federal budget on the basis of the agreement No 075-02-2021-1748.


\begin{thebibliography}{0}
\bibitem{Planck2018}
R.~Adam {\it et al.} [Planck Collaboration],
  \emph{Planck 2018 results. I. Overview and the cosmological legacy of Planck}
  \textit{Astron.\ Astrophys.}\  {\bf 641} (2020) A1
\bibitem{gorb} D.~S.~Gorbunov, V.~A.~Rubakov \emph{Introduction to the early Universe theory: Cosmological perturbations. Inflationary theory} — M.: KRASAND, 2010.
\bibitem{Sch} Maximilian Schlosshauer \emph{Decoherence and the quantum-to-classical transition}, Springer-Verlag, Berlin Heidelberg, 2007 
\bibitem{dec1}
D.~Polarski and A.~A.~Starobinsky,
  \emph{Semiclassicality and decoherence of cosmological perturbations},
  \textit{Class.\ Quant.\ Grav.}\  {\bf 13} (1996) 377
  [gr-qc/9504030].
\bibitem{dec2}
J.~Lesgourgues, D.~Polarski and A.~A.~Starobinsky,
  \emph{Quantum to classical transition of cosmological perturbations for nonvacuum initial states},
  Nucl.\ Phys.\ B {\bf 497} (1997) 479
  [gr-qc/9611019].
\bibitem{dec3}
C.~Kiefer, I.~Lohmar, D.~Polarski and A.~A.~Starobinsky,
  \emph{Pointer states for primordial fluctuations in inflationary cosmology},
  \textit{Class.\ Quant.\ Grav.}\  {\bf 24} (2007) 1699
  [astro-ph/0610700].
  \bibitem{kief} Claus Kiefer \emph{Quantum Gravity}, Oxford University Press, 2007
\bibitem{qc1}
  J.~J.~Halliwell,
  \emph{Decoherence in Quantum Cosmology},
  \textit{Phys.\ Rev.\ D} {\bf 39} (1989) 2912.
\bibitem{qc2}
C.~Kiefer,
  \emph{Continuous Measurement of Minisuperspace Variables by Higher Multipoles},
  \textit{Class.\ Quant.\ Grav.}\  {\bf 4} (1987) 1369.
\bibitem{qc3}
  C.~Kiefer,
  \emph{Decoherence in quantum electrodynamics and quantum gravity},
  \textit{Phys.\ Rev.\ D} {\bf 46} (1992) 1658.
\bibitem{qc4}
A.~O.~Barvinsky, A.~Y.~Kamenshchik, C.~Kiefer and I.~V.~Mishakov,
  \emph{Decoherence in quantum cosmology at the onset of inflation},
  Nucl.\ Phys.\ B {\bf 551} (1999) 374
  [gr-qc/9812043].
\bibitem{qc5}
  C.~Kiefer and D.~Polarski,
  \emph{Why do cosmological perturbations look classical to us?},
  \textit{Adv.\ Sci.\ Lett.}\  {\bf 2} (2009) 164
  [arXiv:0810.0087 [astro-ph]].
\bibitem{BOKam}A.~Y.~Kamenshchik, A.~Tronconi and G.~Venturi,
  \emph{The Born–Oppenheimer method, quantum gravity and matter},
  \textit{Class.\ Quant.\ Grav.}\  {\bf 35} (2018) no.1,  015012
  [arXiv:1709.10361 [gr-qc]].
\bibitem{BOKam1}A.~Y.~Kamenshchik, A.~Tronconi and G.~Venturi,
  \emph{The Born–Oppenheimer approach to quantum cosmology},
  \textit{Class.\ Quant.\ Grav.}\  {\bf 38} (2021) ,  15501
\bibitem{Brout1}
  R.~Brout,
  \emph{On the Concept of Time and the Origin of the Cosmological Temperature},
  \textit{Found.\ Phys.}\  {\bf 17} (1987) 603.\\
\bibitem{Brout2}
R.~Brout and G.~Venturi,
  \emph{Time in Semiclassical Gravity},
  \textit{Phys.\ Rev.\ D} {\bf 39} (1989) 2436.\\
\bibitem{B} Daniel Baumann \emph{TASI Lectures on Inflation}, 2012, [arXiv:0907.5424v2 [hep-th]]
\bibitem{Qg} David Brizuela, Claus Kiefer, Manuel Kraemer \emph{Quantum-gravitational effects on gauge-invariant scalar and tensor perturbations during inflation: The de Sitter case}, \textit{Phys. Rev.} D \textbf{93}, 104035 (2016) [arXiv:1511.05545v2 [gr-qc]] 
\bibitem{Muh} V. Mukhanov, S. Winitzki \emph{Introduction to quantum effects in gravity}, Cambridge, University Press, (2007)
\bibitem{deruelle}N. Deruelle, C. Gundlach, and D. Polarski, \emph{Class. Quantum Grav.} 9, 137 (1992).
\bibitem{martin} J. Martin, V. Vennin, and P. Peter, \textit{Phys. Rev.} D \textbf{86}, 103524 (2012).
\bibitem{NIST}
   [DLMF] NIST Digital Library of Mathematical Functions. http://dlmf.nist.gov/, Release 1.0.14 of 2016-12-21. F. W. J. Olver, A. B. Olde Daalhuis, D. W. Lozier, B. I. Schneider, R. F. Boisvert, C. W. Clark, B. R. Miller, and B. V. Saunders, eds.
\bibitem{Novikov:2017dsl}
O.~O.~Novikov,
Inhomogeneous modes in the PT-symmetric quantum cosmology,
Phys. Part. Nucl. Lett. \textbf{15}, no.4, 353-356 (2018)
\end{thebibliography}
\end{document}